\documentclass[conference]{IEEEtran}

\usepackage{cite}
\usepackage{amsmath,amssymb,amsfonts}
\usepackage{algorithmic}
\usepackage{graphicx}
\usepackage{textcomp}
\usepackage{xcolor}
\usepackage{url}

\usepackage{float}
\usepackage[colorlinks=true,
            linkcolor=blue,
            urlcolor=blue,
            citecolor=blue]{hyperref}

\usepackage{multirow}

\begin{document}




\def\BibTeX{{\rm B\kern-.05em{\sc i\kern-.025em b}\kern-.08em
    T\kern-.1667em\lower.7ex\hbox{E}\kern-.125emX}}

\title{Multimodal Benchmarking and Recommendation of Text-to-Image Generation Models}

\author{\IEEEauthorblockN{Kapil Wanaskar}
\IEEEauthorblockA{\textit{Computer Engineering Dept.} \\
\textit{San José State University}\\
San Jose, CA \\
kapil.wanaskar@sjsu.edu}
\and
\IEEEauthorblockN{Gaytri Jena}
\IEEEauthorblockA{\textit{Independent Researcher} \\
San Jose, CA \\
gaytrijena2000@gmail.com}
\and
\IEEEauthorblockN{Magdalini Eirinaki}
\IEEEauthorblockA{\textit{Computer Engineering Dept.} \\
\textit{San José State University}\\
San Jose, CA \\
magdalini.eirinaki@sjsu.edu}
}

\maketitle

\begin{figure*}[!t]
    \centering
\includegraphics[width=\linewidth]{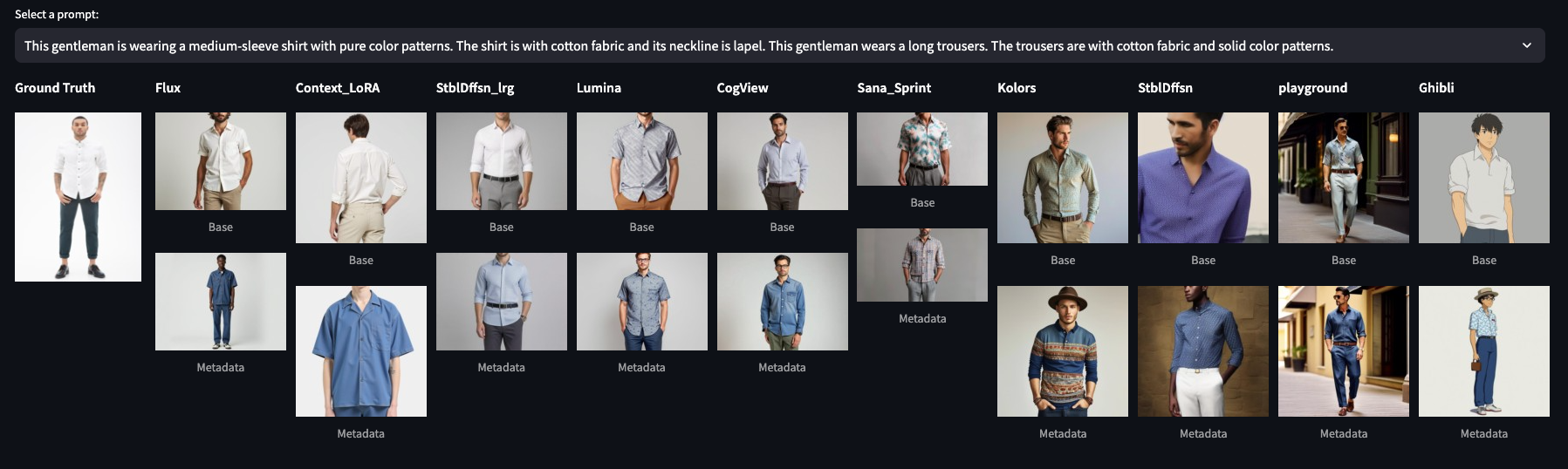}
   \caption{Visual comparison of generated images across models for Prompt 1 (Base vs Metadata)}
   \label{fig:qualitative_prompt1}
\end{figure*}

\begin{figure*}[!t]
    \centering \includegraphics[width=\linewidth]{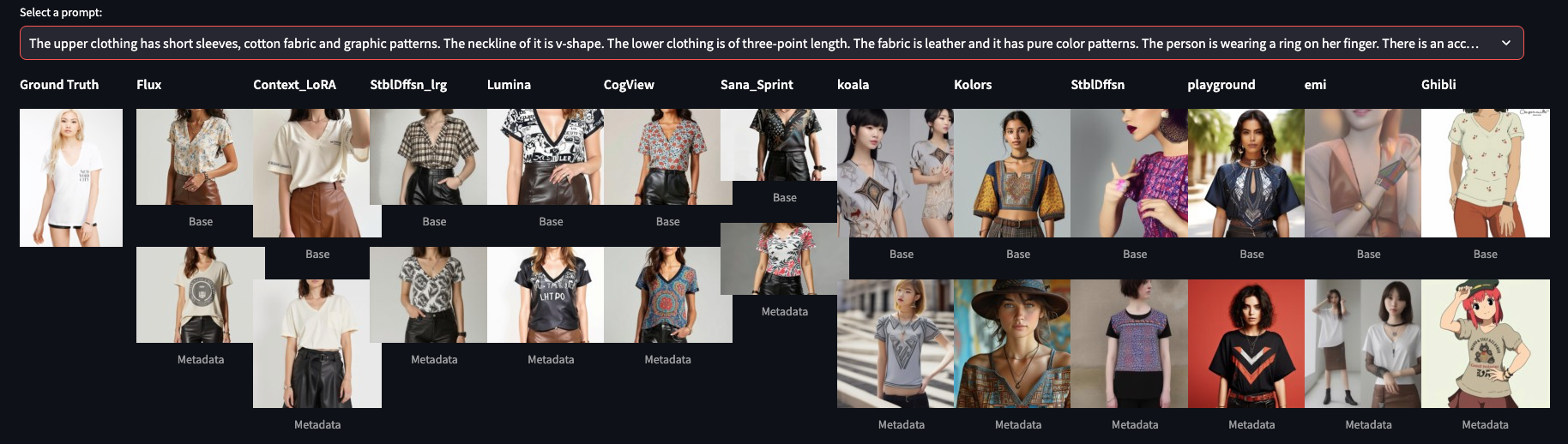}
    \caption{Visual comparison of generated images across models for Prompt 2 (Base vs Metadata)}    \label{fig:qualitative_prompt2}
\end{figure*}

\begin{abstract}
This work presents an open-source unified benchmarking and evaluation framework for text-to-image generation models, with a particular focus on the impact of metadata-augmented prompts. Leveraging the DeepFashion-MultiModal dataset, we assess generated outputs through a comprehensive set of quantitative metrics—including Weighted Score, CLIP (Contrastive Language Image Pre-training)-based similarity, LPIPS (Learned Perceptual Image Patch Similarity), FID (Fr´echet Inception
Distance), and retrieval-based measures—as well as qualitative analysis. Our results demonstrate that structured metadata enrichments greatly enhance visual realism, semantic fidelity, and model robustness across diverse text-to-image architectures. While not a traditional recommender system, our framework enables task-specific recommendations for model selection and prompt design based on evaluation metrics.
\end{abstract}


\section{Introduction}
The success of recent text-to-image (T2I) models has greatly improved the realism and controllability of generated images. However, benchmarking multiple models consistently remains challenging, especially in visually complex domains like clothing synthesis. Existing evaluation pipelines often overlook structured semantic information, limiting the depth of quality assessment.

To address these gaps, we propose a reproducible framework that evaluates T2I models under uniform conditions, combining both quantitative and qualitative analyses. Specifically, we explore the role of metadata-augmented prompts—where structured annotations like sleeve length, neckline, fabric type, and accessories complement the natural language description—to improve generation quality.

Our main contributions are: (1) a unified evaluation pipeline integrating CLIP alignment, LPIPS, FID, MRR, and Recall@K; (2) a large-scale experimental comparison of 10+ publicly available T2I models on the DeepFashion-MultiModal dataset; (3) evidence that metadata-driven prompting enhances semantic fidelity and visual realism; and (4) an interactive, open-source Streamlit for model recommendation and exploration. 
The source code\footnote{\url{https://github.com/kapilw25/Evaluation_generated_images}} of the benchmarking framework as well as a demo app\footnote{\url{https://text2image-bykapil.streamlit.app/}} have been made available to the research community. 

The rest of the paper is organized as follows: in Section II we present a survey of the T2I related work, with supplemental diagrams explaining the pipelines followed by different types of models; in Section III we provide details about the dataset used and in Sections IV and V we discuss our experimental evaluation methodology and results. Finally, we conclude with a discussion and our plans for future work. 



\section{Related Work}

To support future extensibility and systematic benchmarking, we propose a taxonomy-based clustering framework for generative models. Table~\ref{tab:clustering} defines high-level and sub-clusters for organizing models based on their backbone and architectural innovations.

\begin{table*}[!t]
\centering
\caption{Checklist taxonomy for text to image model clustering referring to their architecture}
\label{tab:clustering}
\begin{tabular}{|l|l|l|}
\hline
\textbf{Top-Level Cluster} & \textbf{Sub-Cluster} & \textbf{Example Models} \\
\hline
\multirow{3}{*}{\textbf{Diffusion}} 
& Latent Diffusion & LDM~\cite{rombach2022high}, SDXL~\cite{podell2023sdxl} \\
\cline{2-3}
& Multi-stage Diffusion & CogView3~\cite{ding2022cogview3} \\
\cline{2-3}
& Gated/Group Diffusion & PixArt-$\alpha$~\cite{chen2023pixart}, GDT~\cite{wang2024gdt} \\
\hline
\multirow{1}{*}{\textbf{Transformer-based}} 
& DiT-based & Rectified Flow~\cite{esser2024scaling}, Lumina~\cite{qin2025lumina} \\
\hline
\multirow{2}{*}{\textbf{Hybrid}} 
& GAN + Diffusion & SANA-Sprint~\cite{chen2025sana} \\
\cline{2-3}
& Distillation + Diffusion & KOALA~\cite{lee2023koala} \\
\hline
\end{tabular}
\end{table*}

\subsection{Diffusion}
Diffusion-based models synthesize images through a denoising process from noise in either latent or pixel space. They vary in stage design and attention mechanisms.

\subsubsection{Latent Diffusion}
\begin{figure*}[!t]
    \centering
    \includegraphics[width=0.7\linewidth]{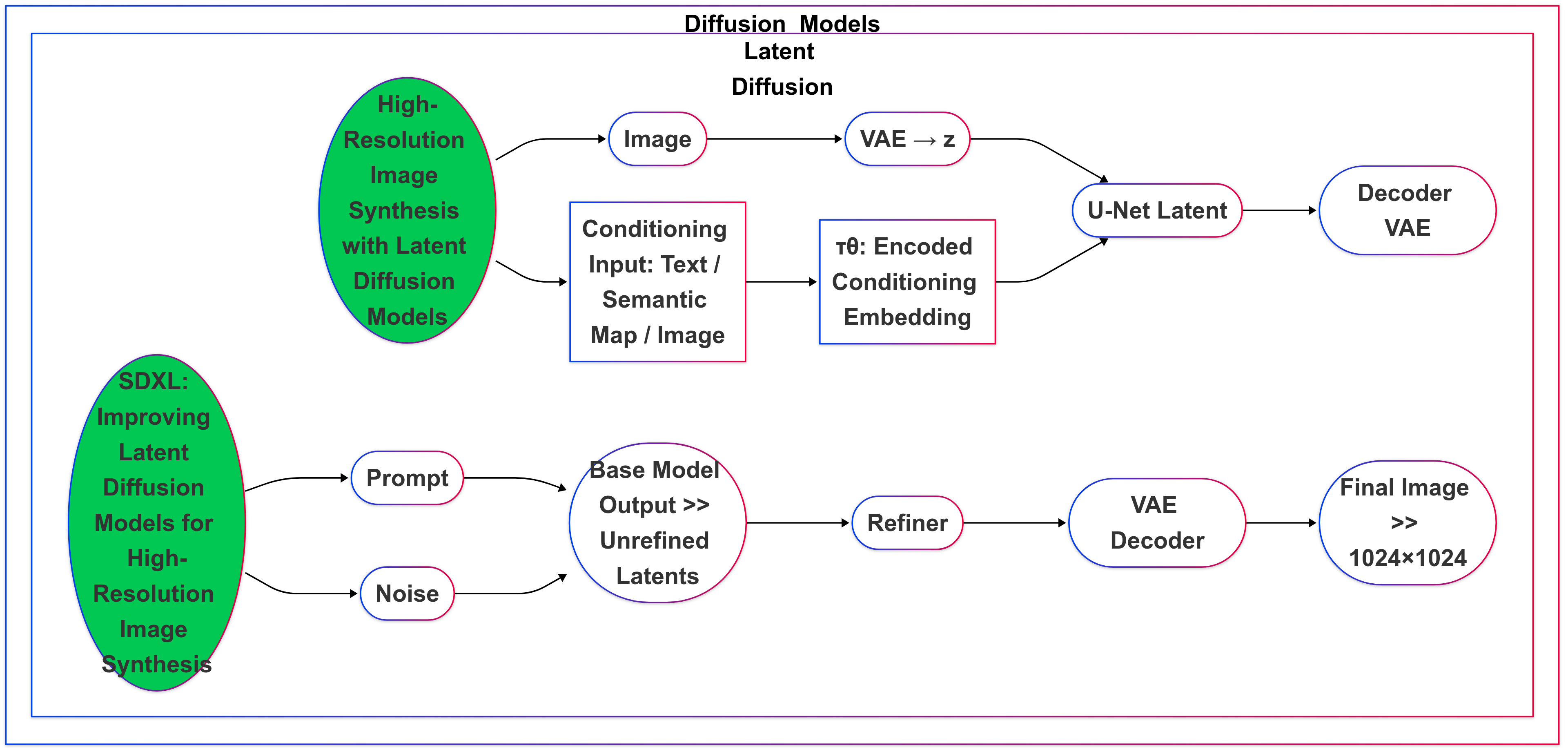}
   \vspace{-0.4cm}
    \caption{Architecture diagrams for Latent Diffusion Models: LDM and SDXL}
    \label{fig:latent_diffusion}
\end{figure*}

Latent diffusion models perform denoising in a compact latent space instead of pixel space, achieving efficiency without sacrificing fidelity. LDM~\cite{rombach2022high} uses a two-stage process: images are encoded via a pretrained VAE, denoised in latent space, and then decoded back, allowing high-resolution synthesis with optional conditioning (text, visual, semantic). SDXL~\cite{podell2023sdxl} extends LDM with a base model producing coarse outputs and a refinement model for enhanced detail, leveraging richer conditioning (e.g., prompts with noise) and generating up to 1024×1024 resolution with improved realism and semantic alignment (see Fig.~\ref{fig:latent_diffusion}).

\subsubsection{Multi-stage Diffusion}
\begin{figure*}[!t]
    \centering
    \includegraphics[width=0.7\linewidth]{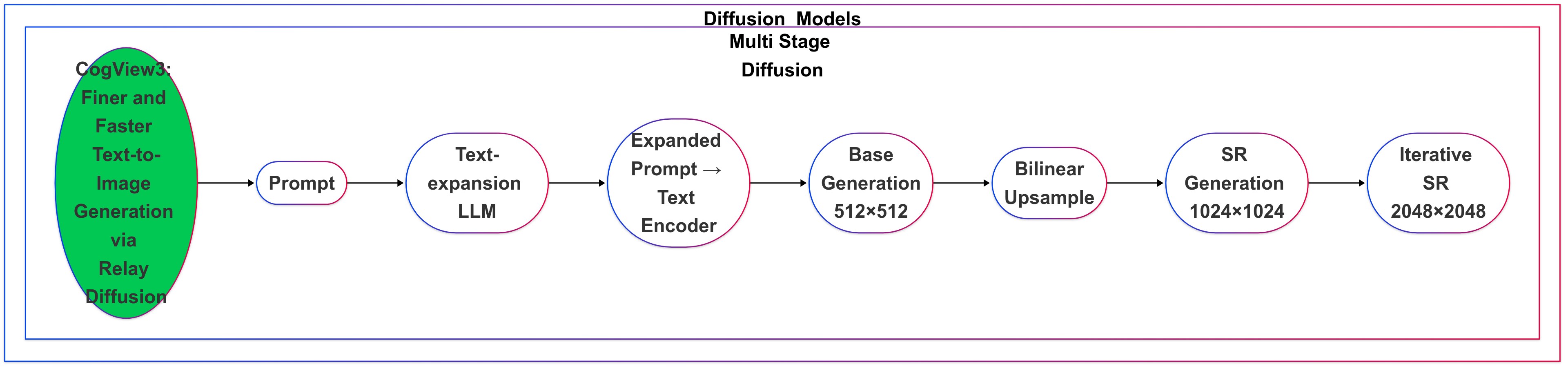}
    \vspace{-0.4cm}
    \caption{Multi-stage diffusion pipeline in CogView3~\cite{ding2022cogview3} using text expansion and relay upscaling}
    \label{fig:multi_stage_diffusion}
\end{figure*}

Multi-stage diffusion models generate images through sequential steps—starting with a low-resolution base and refining it via super-resolution stages to enhance photorealism and detail consistency. CogView3~\cite{ding2022cogview3} exemplifies this with a relay diffusion pipeline: it enriches prompts using a text-expansion language model (LLM), encodes the expanded text, and conditions a diffusion model to produce a 512×512 base image. This is upsampled and refined in stages up to 1024×1024 or 2048×2048 resolution, enabling better structure preservation and fine-grained rendering than single-pass methods (see Fig.~\ref{fig:multi_stage_diffusion}).

\subsubsection{Gated/Group Diffusion}

\begin{figure*}[!t]
    \centering
    \includegraphics[width=0.7\linewidth]{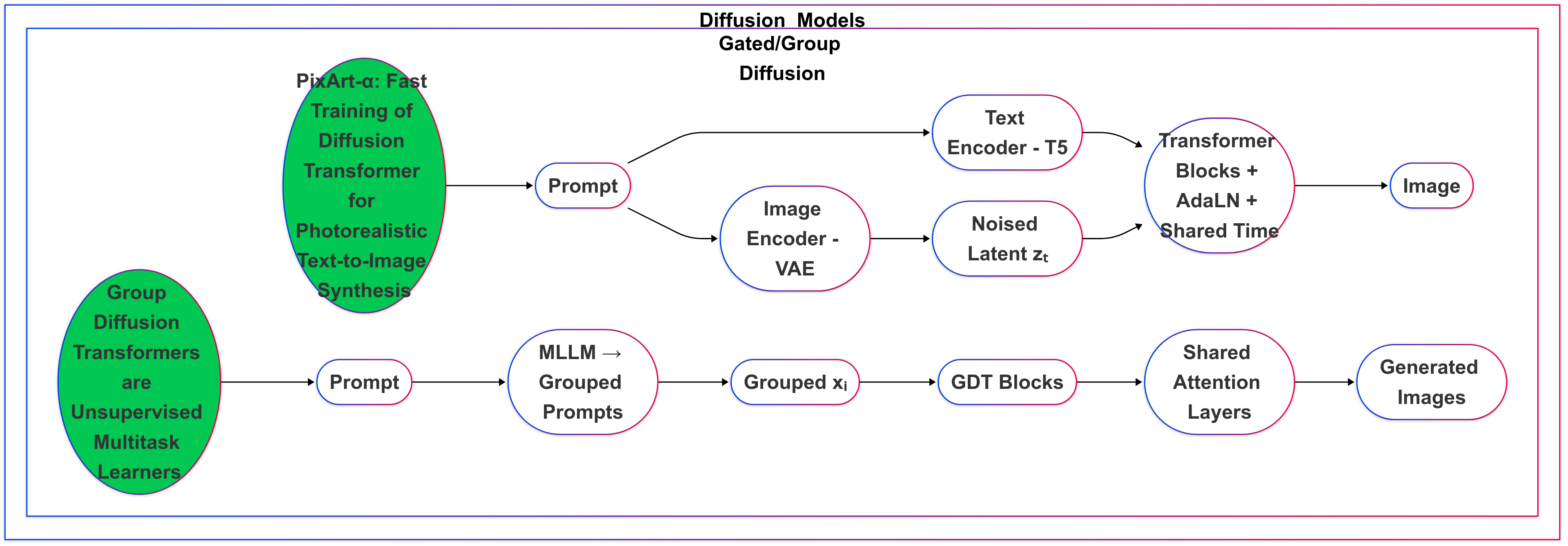}
   \vspace{-0.4cm}
    \caption{Architectures of PixArt-$\alpha$~\cite{chen2023pixart} and GDT~\cite{wang2024gdt} within the Gated/Group Diffusion cluster. PixArt uses transformer blocks with AdaLN; GDT leverages shared attention over grouped prompts.}
    \label{fig:gated_group_diffusion}
\end{figure*}

Gated/Group diffusion models enhance generation through group attention or gated normalization, improving representation learning in complex multi-object scenes. PixArt-$\alpha$~\cite{chen2023pixart} combines T5-encoded text with VAE image latents via transformer blocks featuring multi-head attention and AdaLN, enabling photorealistic synthesis at low training cost. GDT~\cite{wang2024gdt} extends this by decomposing prompts using a multi-modal LLM and applying Group-Attention Blocks to facilitate inter-object learning across grouped images, supporting multitask generation in a single pass (see Fig.~\ref{fig:gated_group_diffusion}).

\subsection{Transformer-based}
Transformer-based models directly apply attention-based blocks across input modalities, typically in DiT-style encoders or dual-stream designs.

\subsubsection{DiT-based}
\begin{figure*}[!t]
    \centering
    \includegraphics[width=0.55\linewidth]{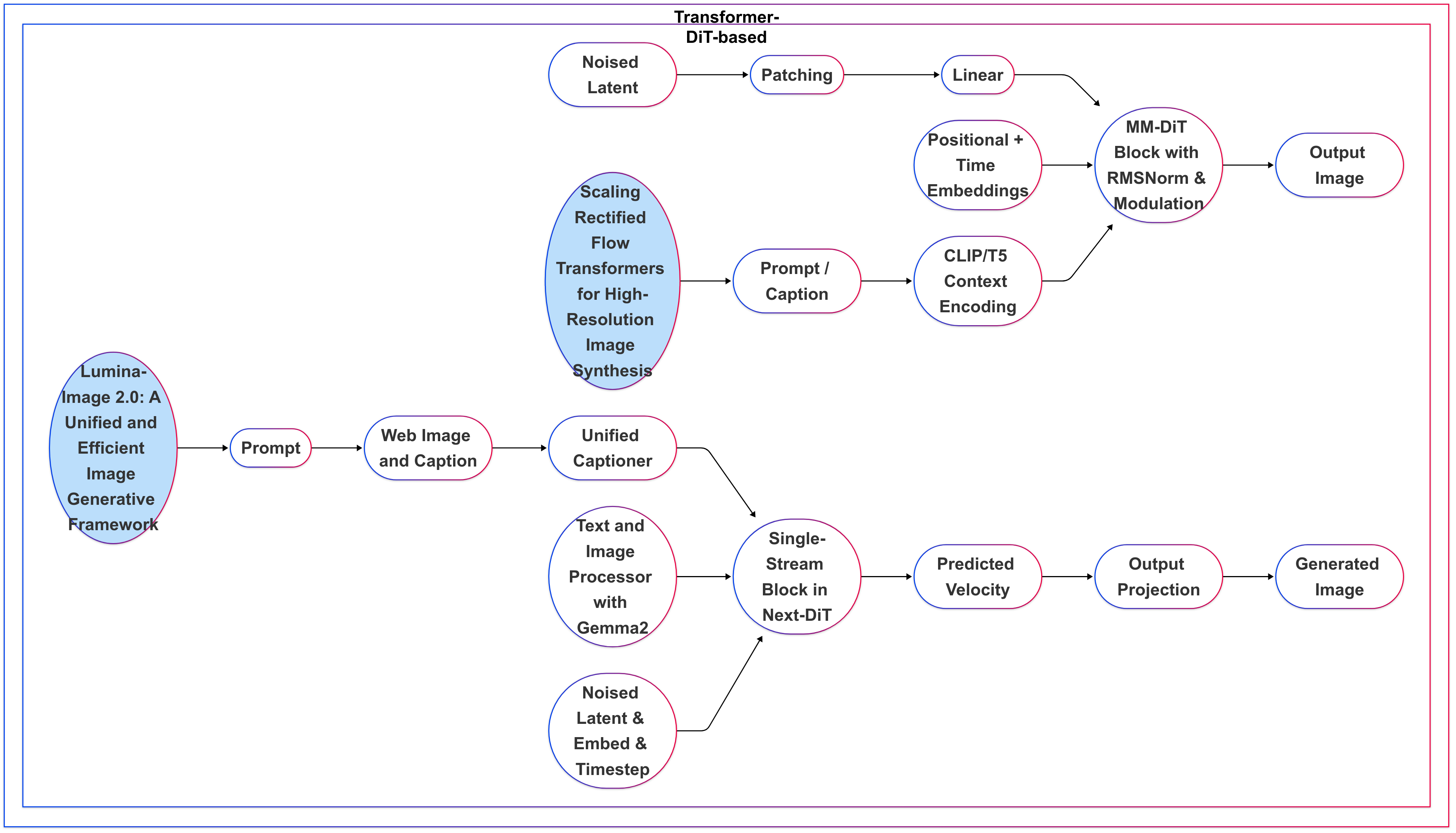}
   \vspace{-0.4cm}
    \caption{Architectures of Rectified Flow~\cite{esser2024scaling} and Lumina~\cite{qin2025lumina} within the DiT-based cluster. Rectified Flow adopts MM-DiT with modality-specific normalization; Lumina combines unified captioning and single-stream DiT-based generation.}
    \label{fig:dit_based_transformers}
\end{figure*}

DiT-based models replace U-Net with transformer-based denoising networks, offering better modularity for high-resolution generation. Rectified Flow~\cite{esser2024scaling} uses a rectified diffusion process and Multi-Modality DiT (MM-DiT) blocks that integrate linear projections, time embeddings, and CLIP/T5 conditioning, with RMSNorm and conditional modulation for stable, aligned outputs. Lumina~\cite{qin2025lumina} employs a unified captioning system and single-stream DiT generator, processing noised latents and hierarchical captions through Gemma2 encoders and predicting velocity vectors for final image reconstruction, achieving efficient and domain-general generation (see Fig.~\ref{fig:dit_based_transformers}).

\begin{figure*}[!t]
    \centering
    \includegraphics[width=0.55\linewidth]{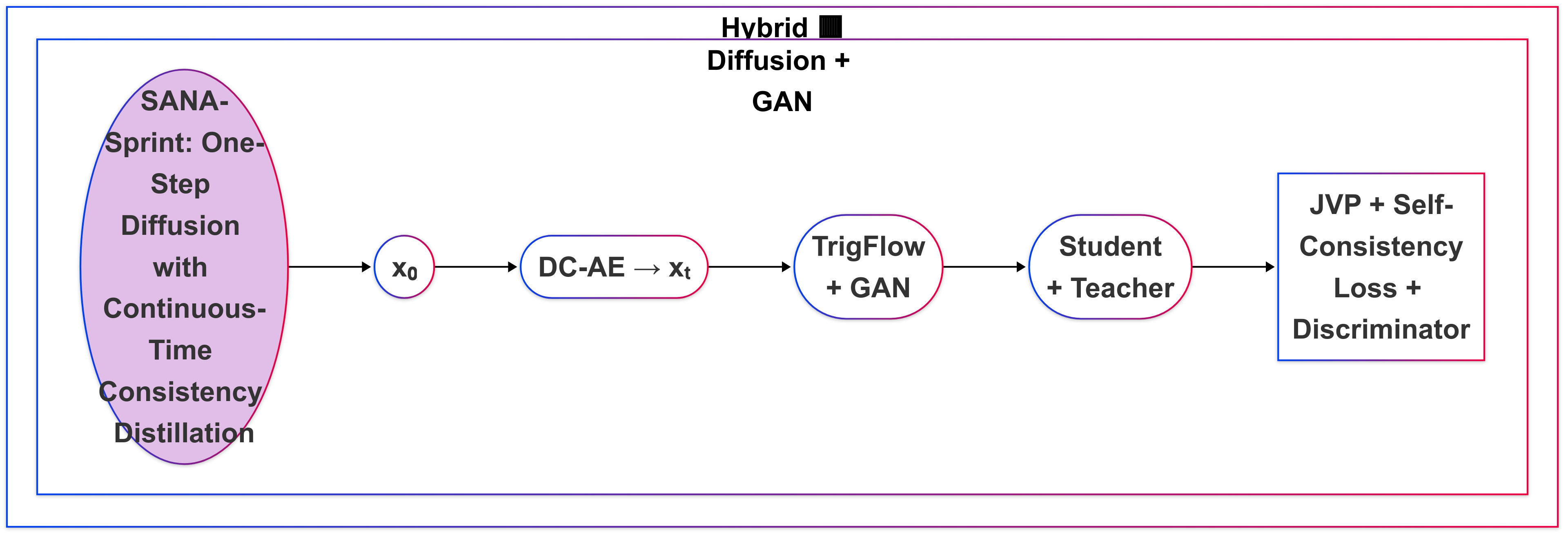}
  \vspace{-0.4cm}
    \caption{Architecture of SANA-Sprint~\cite{chen2025sana}, which combines TrigFlow-conditioned diffusion with GAN-based consistency and adversarial losses. A DC-AE encoder reduces input to latent form before one-step prediction.}
    \label{fig:gan_diffusion_sana}
\end{figure*}

\subsection{Hybrid}
Hybrid models blend components from different paradigms—typically combining GANs, diffusion, and distillation for faster or more stable training.

\subsubsection{GAN + Diffusion}
This hybrid paradigm blends consistency-driven diffusion generation with adversarial training from GANs. These models leverage both the semantic alignment strength of diffusion and the realism boost from discriminative feedback.

\paragraph{SANA-Sprint~\cite{chen2025sana}}  
SANA-Sprint proposes a one-step generation pipeline combining TrigFlow-based time conditioning with adversarial GAN supervision. The model first encodes a clean image $x_0$ into a latent $x_t$ using a DC-Autoencoder (DC-AE). Then, a TrigFlow transformation modulates this latent with sine and cosine temporal embeddings, feeding it to a shared Student–Teacher network. The Teacher generates target velocities and features, while the Student predicts reconstructed outputs. Both JVP-based self-consistency loss and adversarial loss from GAN-based discriminators are used. This training allows high-quality image synthesis from a single forward pass in latent space (see Fig.~\ref{fig:gan_diffusion_sana}).

\subsubsection{Distillation + Diffusion}
Distillation-based models aim to reduce diffusion steps or compress large models via teacher-student learning. KOALA ~\cite{lee2023koala} adopts this approach.

\section{Dataset}
\begin{figure*}[!t]
    \centering
    \includegraphics[width=0.8\linewidth]{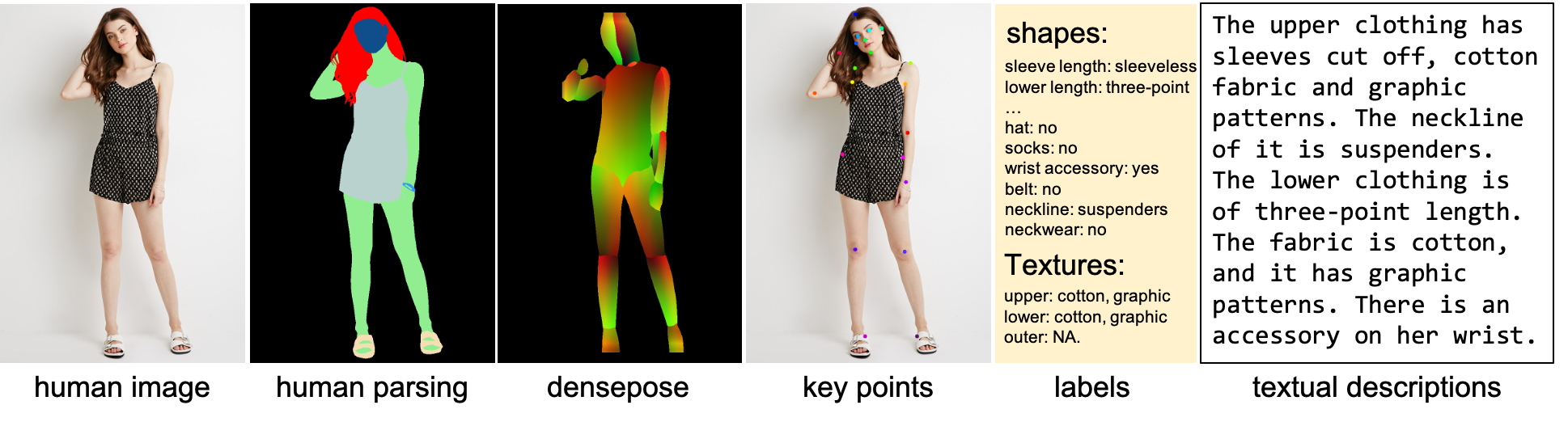}
    \vspace{-0.4cm}
    \caption{DeepFashion-MultiModal dataset \cite{jiang2022text2human}: Left to right — human image, human parsing, DensePose, keypoints, structured labels, and textual descriptions.}
    \label{fig:dataset_pipeline}
\end{figure*}

We employ the DeepFashion-MultiModal dataset \cite{jiang2022text2human}, which contains 44,096 high-resolution human images with rich annotations including parsing masks (24 classes), DensePose outputs, 21 annotated keypoints, shape/fabric/color labels, and manually curated captions. For our experiments, textual descriptions serve as base prompts, while shape, fabric, and color labels are used to construct metadata-augmented prompts. The dataset provides fine-grained control over attributes like sleeve length, neckline, fabric type, and accessories (see Fig.~\ref{fig:dataset_pipeline}). All annotations were processed via \texttt{metadata\_preprocessing.py}, producing CSV mappings of base and enriched prompts for consistent use across generation and evaluation.

\section{Experimental Setup}

\begin{figure*}[!t]
    \centering
    \includegraphics[width=0.9\textwidth, height=0.85\textheight, keepaspectratio, trim=0.3cm 0.3cm 0.3cm 0.3cm, clip]{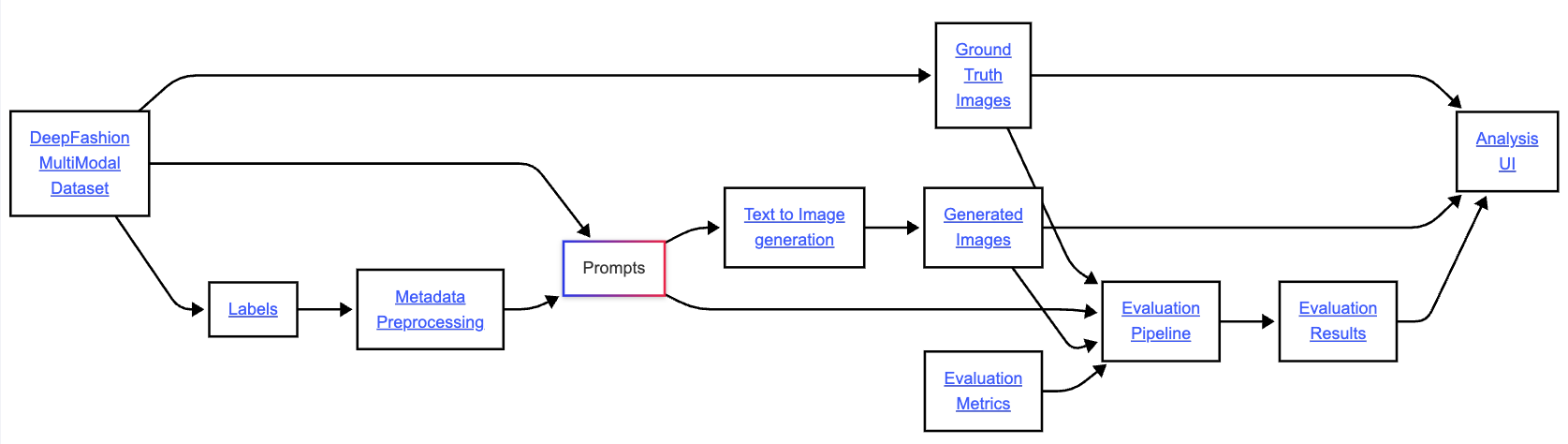}
    \caption{Overall system architecture for text-to-image evaluation and recommendation}
    \label{fig:architecture}
\end{figure*}

Each model is tested on two prompt types: (1) base prompts from DeepFashion captions and (2) metadata-enriched prompts including garment attributes. This enables comparison with and without semantic cues. Evaluation covers 10+ state-of-the-art models—Stable Diffusion, CogView, Flux, Context-LoRA, Lumina—selected for architectural variety: diffusion, autoregressive, and LoRA-enhanced types.  (see Fig.~\ref{fig:architecture})

For our evaluation we use different metrics used to evaluate image generation and similarity: CLIP (Contrastive Language-Image Pre-training) is a model that provides similarity scores between images and text descriptions, useful for image captioning and retrieval; LPIPS (Learned Perceptual Image Patch Similarity) measures the perceptual similarity between images, aiming to capture human perception; FID (Fréchet Inception Distance) assesses the similarity between distributions of features extracted from real and generated images, indicating how well generated images resemble real ones. Retrieval scores are metrics used to evaluate the effectiveness of retrieving relevant images from a collection based on a query (in this study we use recall@K). 

The \texttt{evaluation\_pipeline.py} automates benchmarking using \texttt{gen\_img\_metadata.csv} to map each generated image to its ground truth. For each model, multiple metrics are computed: CLIP Score (prompt vs generated image) using CLIP ViT-B/32; CLIP Cosine Similarity (generated vs ground truth); LPIPS for perceptual similarity; FID for distributional gap using InceptionV3 features; MRR for reciprocal rank of ground truth; and Recall@3 for top-3 match accuracy. A composite Weighted Score is used to aggregate normalized metrics:
$0.4N_{\text{CLIP}} + 0.3N_{\text{LPIPS}} + 0.15N_{\text{FID}} + 0.1N_{\text{Ret}} + 0.05N_{\text{CLIP}_{\text{prompt}}}$,
where \(N\) is min–max–scaled (inverted for LPIPS, FID).
All metrics are saved to \texttt{evaluation\_results.csv}. Models are ranked by Weighted Score to identify top performers.

\section{Results}
\vspace{-0.3cm}
\subsection{Quantitative}

\subsubsection{Metric Evaluations}
\begin{figure*}[!t]
    \centering
    \includegraphics[width=0.65\linewidth]{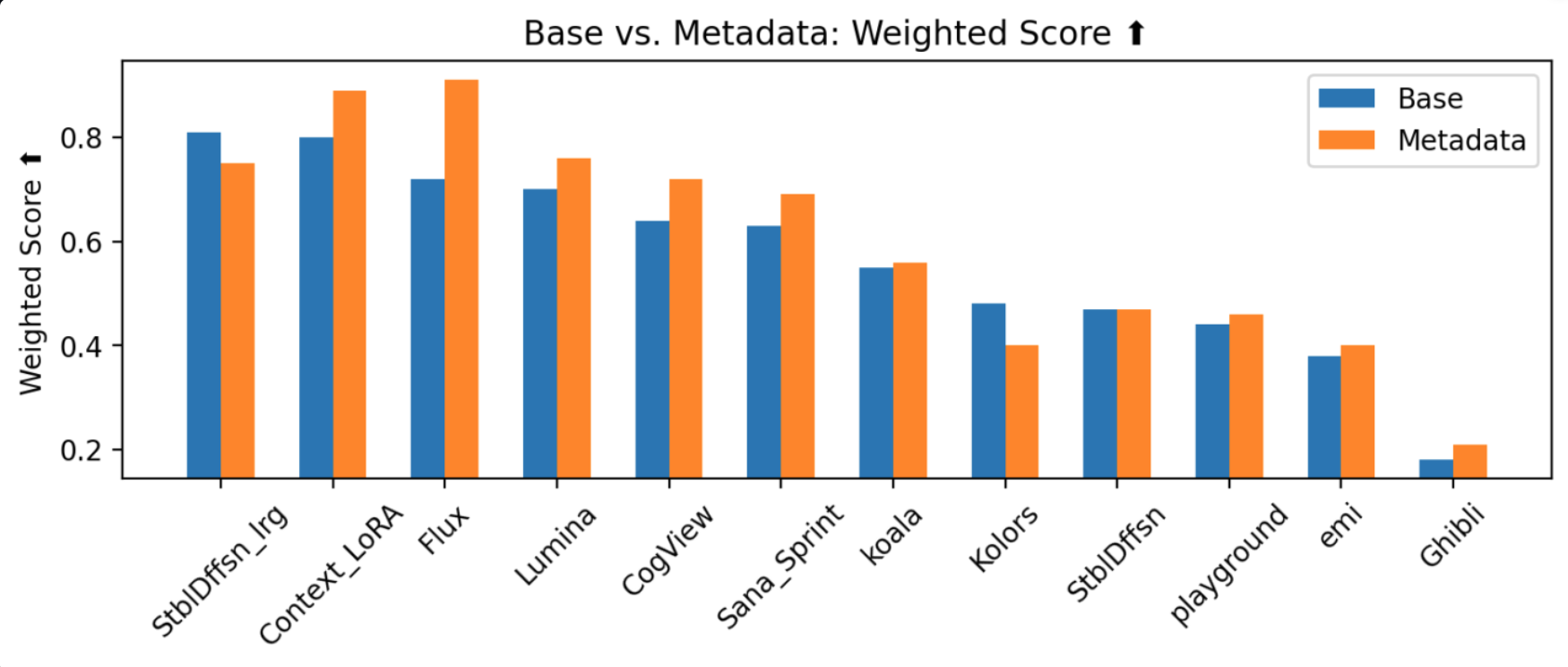}
    \vspace{-0.4cm}
    \caption{Comparison of Weighted Score between Base and Metadata-augmented prompts}
   \label{fig:quantitative_weighted_score}
\end{figure*}

\begin{figure*}[!t]
    \centering
    \includegraphics[width=0.65\linewidth]{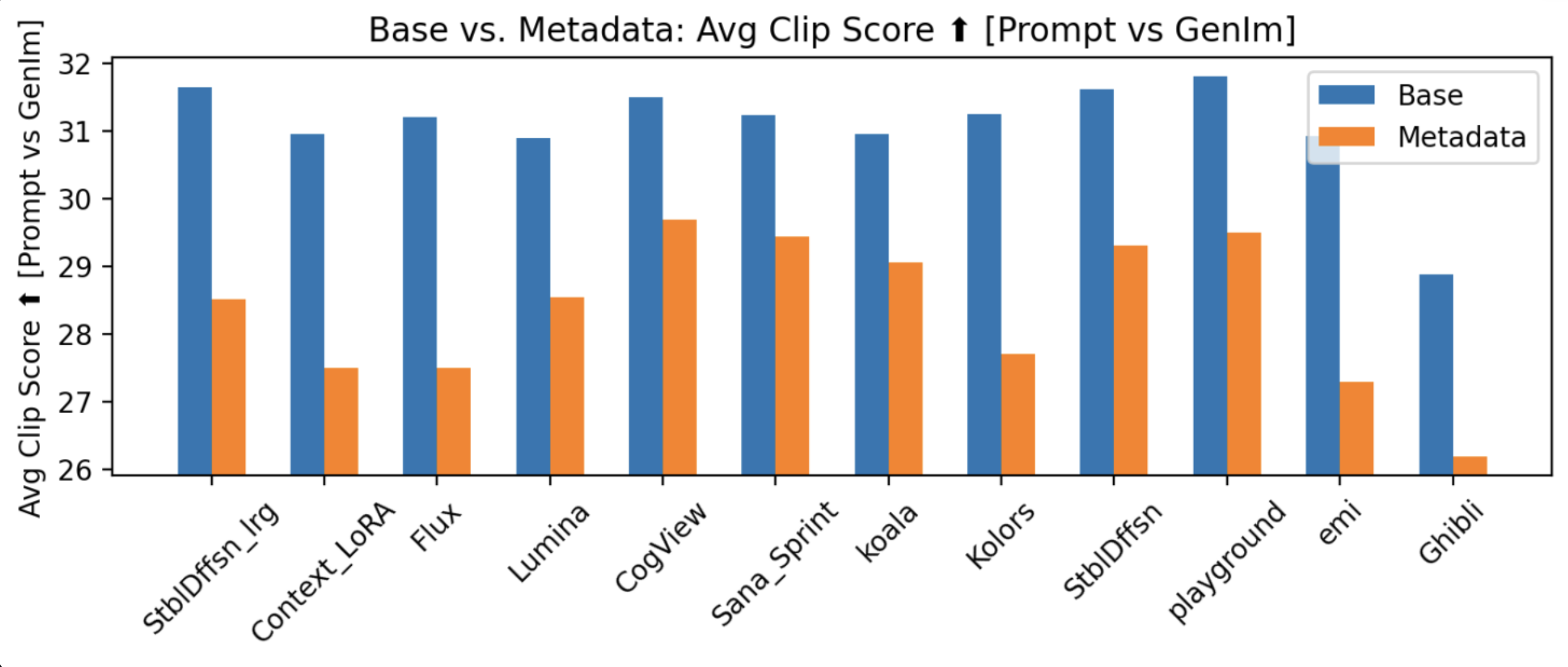}
    \vspace{-0.4cm}
    \caption{Comparison of Avg CLIP Score [Prompt vs Generated Image] between Base and Metadata prompts} \label{fig:quantitative_clip_genimg_prompt}
\end{figure*}

We assessed metadata-augmented prompts across three quantitative metrics: Weighted Score, CLIP Score (Prompt vs Generated Image), and CLIP Cosine Similarity (Generated vs Ground Truth Image). The Weighted Score (Fig.~\ref{fig:quantitative_weighted_score}) aggregates normalized CLIP, LPIPS, FID, and retrieval scores. Metadata-enhanced prompts improved this metric across most models, especially for Flux and Context\_LoRA, indicating that added semantic detail substantially improves generation quality. In contrast, Avg CLIP Score (Prompt vs Generated Image) (Fig.~\ref{fig:quantitative_clip_genimg_prompt}) showed a slight drop with metadata prompts—base prompts performed better in semantic alignment. The added richness occasionally introduced deviation from the minimal prompt intent.

\begin{figure*}[!t]
    \centering
    \includegraphics[width=0.65\linewidth]{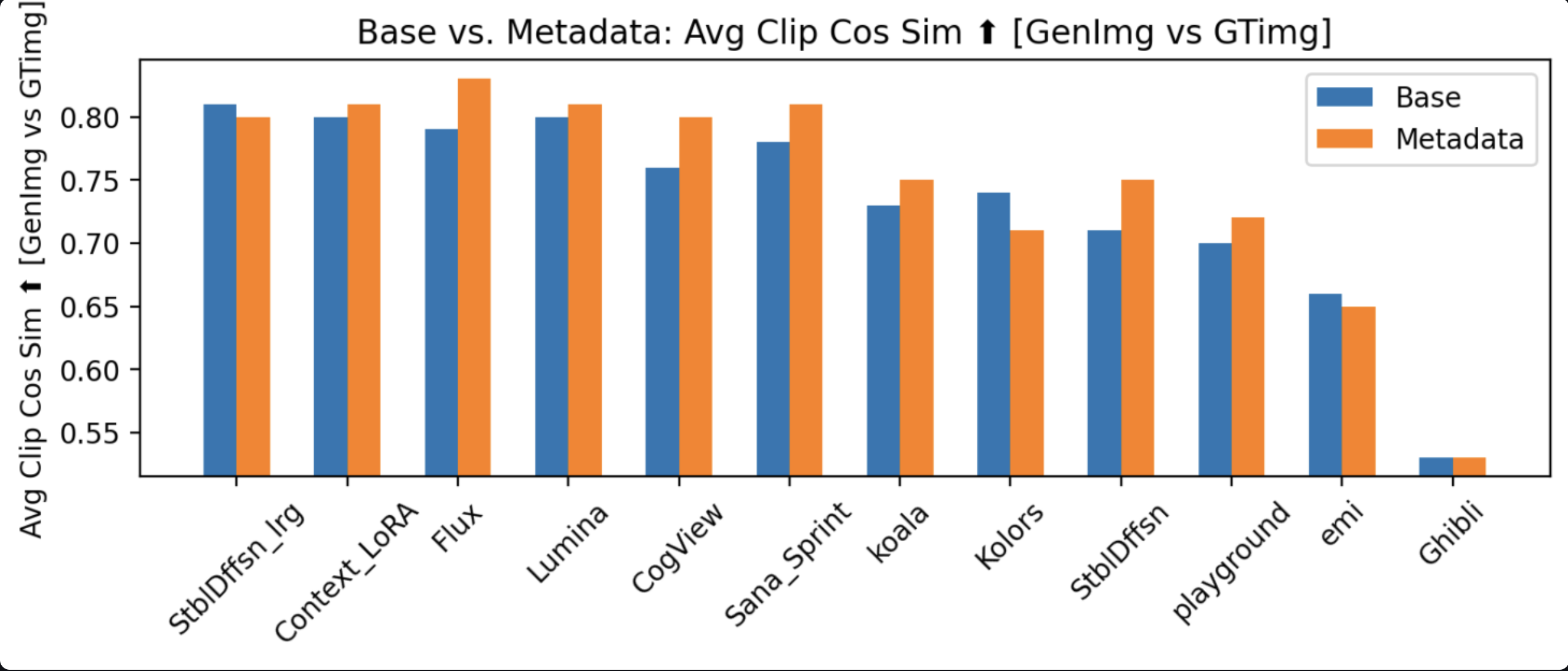}
     \vspace{-0.4cm}
    \caption{Comparison of Avg CLIP Cosine Similarity [Generated Image vs Ground Truth Image]}
    \label{fig:quantitative_clip_cos_genimg_gtimg}
\end{figure*}

Avg CLIP Cosine Similarity (Generated vs Ground Truth) (Fig.~\ref{fig:quantitative_clip_cos_genimg_gtimg}) demonstrated consistent gains for metadata-augmented prompts, reflecting improved alignment with ground-truth images. Models like Flux, Sana-Sprint, and CogView achieved notably higher scores, confirming the benefit of metadata for visual fidelity. Overall, metadata improves realism and ground-truth grounding at a minor cost to prompt-image alignment, favoring realism for most use cases.

\subsubsection{Model Comparisons}

\begin{figure*}[!t]
    \centering
    \includegraphics[width=0.45\linewidth]{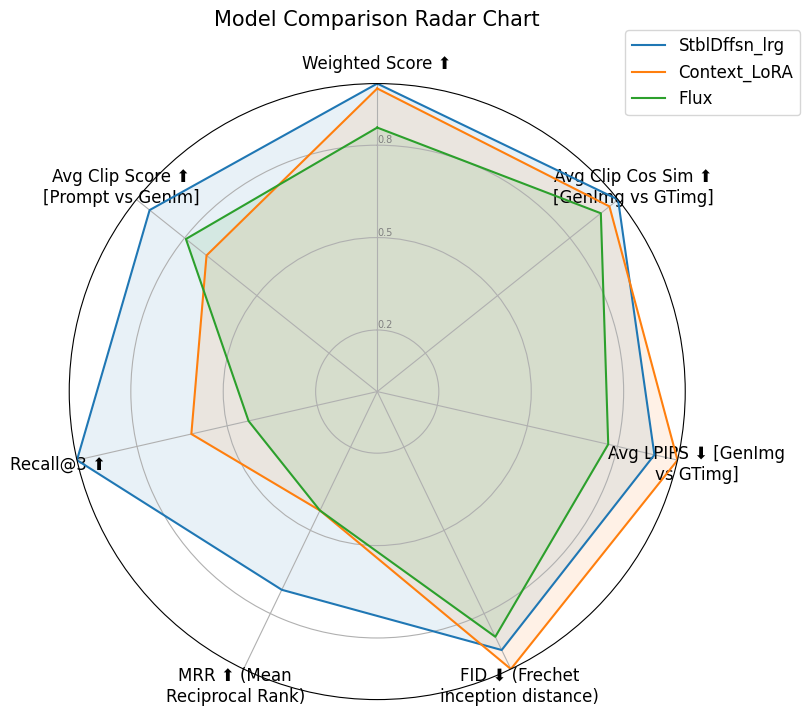}
    \caption{Radar chart comparing Flux, Context\_LoRA, and StblDffsn\_lrg across all metrics}
    \label{fig:quantitative_radar_chart}
\end{figure*}

\textbf{Radar Chart} (Fig.~\ref{fig:quantitative_radar_chart}): This chart compares the Top 3 models - Flux, Context\_LoRA, and StblDffsn\_lrg - across six metrics: Weighted Score, Avg CLIP Cosine Similarity, LPIPS, FID, MRR, and Recall@3. StblDffsn\_lrg leads in Weighted Score and CLIP Score, indicating strong generation quality and prompt fidelity. Context\_LoRA achieves the best FID but scores lower in Recall@3. Flux performs steadily overall but underperforms in LPIPS and MRR. Overall, StblDffsn\_lrg shows the best all-round performance, with Context\_LoRA close behind.

\begin{figure*}[!t]
    \centering
    \includegraphics[width=0.7\linewidth]{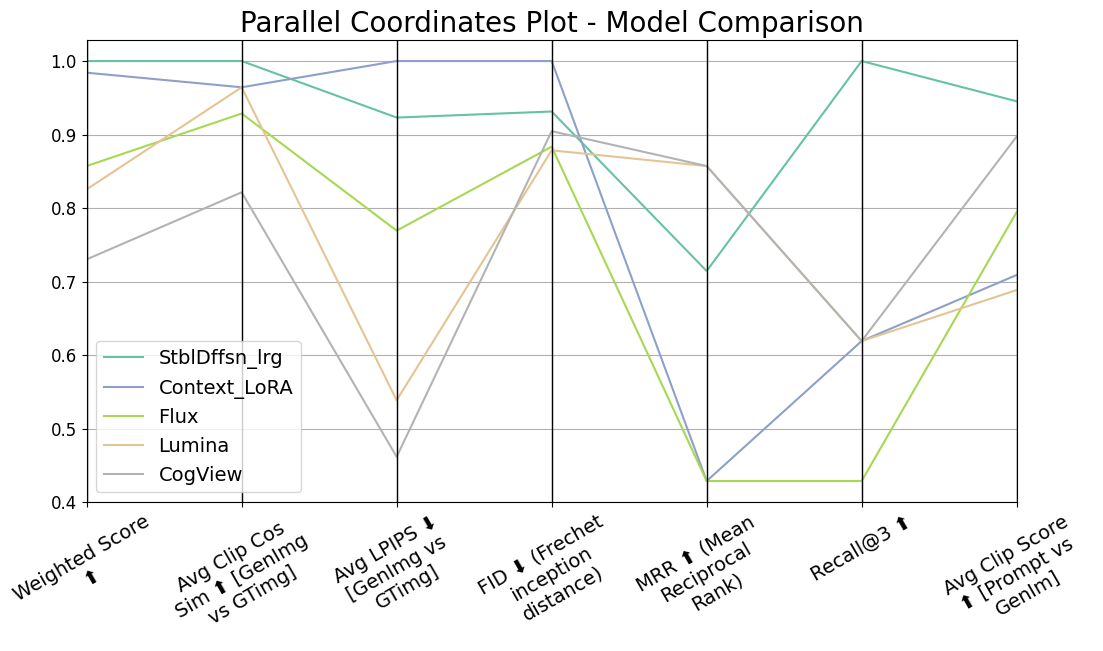}
    \vspace{-0.5cm}
    \caption{Parallel Coordinates Plot comparing Flux, Context\_LoRA, StblDffsn\_lrg, Lumina, and CogView across all metrics}  \label{fig:quantitative_parallel_coordinates}
\end{figure*}

\textbf{Parallel Coordinates Plot} (Fig.~\ref{fig:quantitative_parallel_coordinates}): This plot visualizes the Top 5 models - Flux, Context\_LoRA, StblDffsn\_lrg, Lumina, and CogView - across six metrics. Context\_LoRA maintains consistently strong scores, particularly in CLIP Cosine Similarity and FID. StblDffsn\_lrg excels in Weighted Score and Recall@3, while Flux is strong in alignment but weaker in LPIPS. Lumina and CogView trail slightly but show good CLIP Cosine Similarity and FID. Trade-offs between metrics are apparent, with Context\_LoRA and StblDffsn\_lrg emerging as the most reliable.

\begin{figure*}[!t]
    \centering
       \vspace{-0.7cm}
    \includegraphics[width=0.7\linewidth]{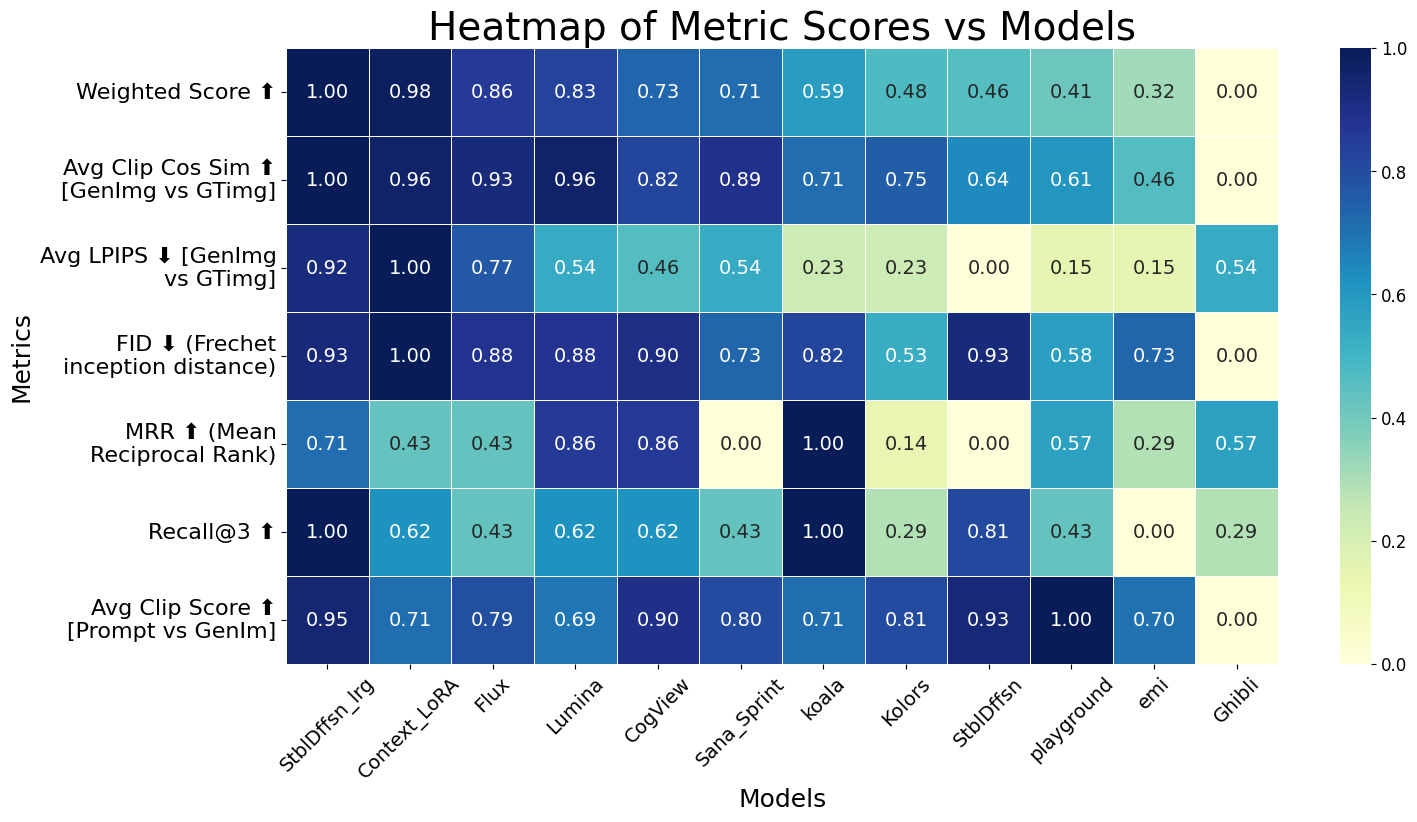}
    \caption{Heatmap showing normalized metric scores for all models}
    \label{fig:quantitative_heatmap}
\end{figure*}

\textbf{Metric Heatmap} (Fig.~\ref{fig:quantitative_heatmap}): The heatmap shows normalized scores across all models. StblDffsn\_lrg and Context\_LoRA dominate in Weighted Score, CLIP Cosine Similarity, and FID. Flux performs well in CLIP-based metrics but moderately in LPIPS and Recall@3. Lumina and CogView offer solid CLIP Cosine and FID performance but slightly weaker perceptual and retrieval scores. Sana\_Sprint and KOALA remain balanced but mid-tier. Ghibli, emi, and Playground underperform, with Ghibli ranking lowest overall.

\begin{figure*}[!t]
    \centering
    \includegraphics[width=0.7\linewidth]{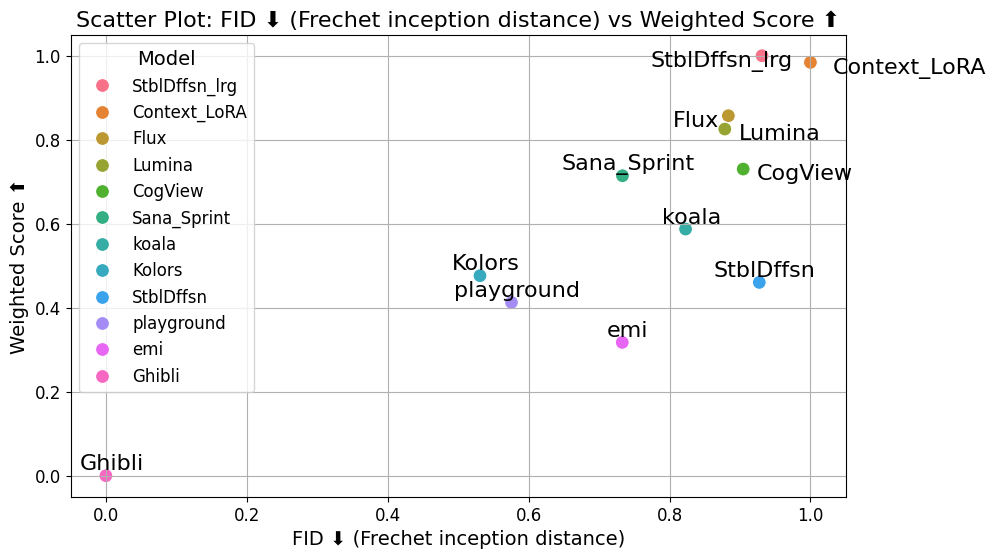}
    \caption{Scatter plot of FID (lower is better) vs Weighted Score (higher is better) across all models}
\label{fig:quantitative_scatter_fid_weightedscore}
\end{figure*}

\textbf{Scatter Plot: FID vs Weighted Score} (Fig.~\ref{fig:quantitative_scatter_fid_weightedscore}): This plot reveals how FID and Weighted Score relate across models. StblDffsn\_lrg and Context\_LoRA score best, balancing low FID and high Weighted Score. Flux and Lumina follow with strong positions. CogView and Sana\_Sprint perform decently but with slightly higher FID. Ghibli is a major outlier — low FID but poor Weighted Score, indicating semantic failure. emi also fares poorly. The visualization confirms top models deliver both quality and alignment, unlike underperformers.

\subsection{Qualitative}

To complement the quantitative results, we conducted a visual comparison of generated images against ground-truths using two prompts — one simple and one complex — evaluated both with and without metadata. In Prompt 1 (Fig.~\ref{fig:qualitative_prompt1}), the base prompt described medium-sleeve cotton shirts with lapel necklines and long cotton trousers. Metadata-enhanced prompts appended structured attributes such as gender, category, sleeve length, neckline, and fabric. Without metadata, models like Flux and Context\_LoRA occasionally misrepresented sleeve lengths or materials. Metadata enrichment improved structural accuracy, particularly for fit, fabric texture, and neckline details. Flux and CogView showed notable improvements, while Kolors and Ghibli benefited from more semantically aligned outputs despite stylistic tendencies.

Prompt 2 (Fig.~\ref{fig:qualitative_prompt2}) involved a more complex scene with accessories and multi-material outfits. The base prompt described short sleeves, v-shaped necklines, leather skirts, and the presence of a ring and wrist accessory. Metadata versions included detailed garment and accessory labels. Without metadata, models often omitted accessories or failed to depict leather textures accurately. Enriched prompts helped models like Sana-Sprint, Flux, and Context\_LoRA generate more faithful outputs, rendering v-necklines, visible rings, and wristbands correctly. Even stylistically unique models like Ghibli showed better semantic alignment with metadata. Overall, metadata-enhanced prompts significantly boosted fine-detail accuracy and scene coherence, especially in complex scenarios.

\section{Conclusion}

This study offers a holistic evaluation of metadata-augmented prompts for text-to-image generation by integrating quantitative metrics and qualitative analysis. Our metric-based assessments—including Weighted Score, Prompt-to-Image CLIP Score, and CLIP Cosine Similarity with ground-truth images—revealed that metadata consistently improved composite image quality and visual realism, albeit with slight trade-offs in prompt fidelity. Model-wise, StblDffsn\_lrg and Context\_LoRA led performance across Radar, Parallel Coordinate, Heatmap, and FID-Weighted Score Scatter plots, showcasing strong trade-offs between realism and retrieval alignment. Qualitative results further validated these findings, with metadata improving fine-grained visual attributes like fabric, sleeve length, and accessory realism across both simple and complex prompts. Notably, models such as Flux, Context\_LoRA, and Sana-Sprint demonstrated enhanced consistency and semantic grounding with enriched prompts. Overall, metadata enrichment proves to be an effective strategy for boosting image fidelity and semantic alignment, while our multi-faceted benchmarking framework offers robust guidance for model selection, evaluation, and future generative research.

\section{Future Work}
Future work will address several promising directions to enhance benchmarking and generation quality. First, to standardize evaluation across all models, we plan to build dedicated inference APIs for high-performing, non-HuggingFace-hosted models such as ``SimianLuo/LCM\_Dreamshaper\_v7'' \cite{luo2023latent} and ``h94/IP-Adapter-FaceID'' \cite{ye2023ipadapter} by replicating their diffusion pipelines and securely hosting them. Second, we aim to fine-tune base diffusion models using metadata-enriched datasets, which could further boost Weighted Score and Ground-Truth Cosine Similarity while preserving prompt-image alignment. Third, AutoML-based prompt tuning will be explored to automate metadata embedding and prompt refinement, potentially improving performance with less manual effort. Fourth, based on our findings of model variance across garment types and accessories, we propose developing a user-personalized generation recommendation system that tailors model selection and prompt metadata to individual user needs. Lastly, we plan to study dynamic metadata injection strategies such as inserting metadata during intermediate denoising steps, optimizing prompts under metadata constraints, and adaptive filtering of metadata attributes to better balance semantic richness and prompt fidelity.

\bibliographystyle{IEEEtran}
\bibliography{references} 

\end{document}